\documentclass[12pt,preprint]{aastex}
\begin{document}
\newcommand{\msun}{M_{\odot}}
\newcommand{\kms}{\, {\rm km\, s}^{-1}}
\newcommand{\cm}{\, {\rm cm}}
\newcommand{\gm}{\, {\rm g}}
\newcommand{\gev}{\, {\rm GeV}}
\newcommand{\erg}{\, {\rm erg}}
\newcommand{\kpc}{\, {\rm kpc}}
\newcommand{\mpc}{\, {\rm Mpc}}
\newcommand{\seg}{\, {\rm s}}
\newcommand{\kev}{\, {\rm keV}}
\newcommand{\hz}{\, {\rm Hz}}
\newcommand{\nhi}{N_{\hi}}
\newcommand{\etal}{et al.\ }
\newcommand{\yr}{\, {\rm yr}}
\newcommand{\eq}{eq.\ }
\def\arcsec{''\hskip-3pt .}

\title{A Test of the Collisional Dark Matter Hypothesis from Cluster
Lensing}
\author{Jordi Miralda-Escud\'e$^{1}$}
\affil{The Ohio State University, Dept. of Astronomy,
McPherson Labs.,
140 W. 18th Ave., Columbus, OH 43210}
\authoremail{jordi@astronomy.ohio-state.edu}
\affil{$^{1}$ Alfred P. Sloan Fellow}

\begin{abstract}

  Spergel \& Steinhardt proposed the possibility that the dark matter
particles are self-interacting, as a solution to two discrepancies
between the predictions of cold dark matter models and the observations:
first, the observed dark matter distribution in some dwarf galaxies has
large, constant-density cores, as opposed to the predicted central
cusps; and second, small satellites of normal galaxies are much less
abundant than predicted. The dark matter self-interaction would produce
isothermal cores in halos and expel the dark matter particles from
dwarfs orbiting in large halos. Another consequence of the model is that
halos should become spherical once most particles have interacted.
Several observations show that the mass distribution in relaxed clusters
of galaxies is elliptical. Here, I discuss in particular gravitational
lensing in the cluster MS2137-23, where the ellipticity of the dark
matter distribution can be measured to a small radius, $r\sim 70$ kpc,
suggesting that most dark matter particles in clusters outside this
radius do not collide during the characteristic age of clusters. If
true, this implies that any dark matter self-interaction with a cross
section independent of velocity is too weak to have affected the
observed density profiles in the dark-matter dominated dwarf galaxies,
or to have facilitated the destruction of dwarf satellites in galactic
halos. If $s_x$ is the cross section and $m_x$ the mass of the dark
matter particle, then $s_x/m_x < 10^{-25.5} \cm^2/\gev$.

\end{abstract}

\keywords{dark matter - galaxies: clusters: general - galaxies: formation
 - large-scale structure of universe}

\section{Introduction}

  The Cold Dark Matter (CDM) model of structure formation in
the universe has been tremendously successful in accounting for a huge
variety of available observations (e.g., the Cosmic Background fluctuations,
the abundances of clusters of galaxies, peculiar velocity fields,
the Ly$\alpha$ forest), provided that the mean density of matter
is only a fraction $ \Omega_m \simeq 0.3$ of the critical density, and
the existence of vacuum energy with a negative pressure equation of
state is allowed to make the universe spatially flat (e.g.,
\cite{kp00}; \cite{ptw99}; \cite{bops99}; \cite{sw95}; \cite{ecf96};
\cite{crofta99}).

  A possible problem of this model has emerged when comparing the
density profiles of dark matter halos predicted in numerical
simulations, with observations of the rotation curves in dwarf galaxies
(\cite{m94}; \cite{fp94}; \cite{nfw96}; \cite{moorea98}; \cite{kkbp98};
\cite{moorea99}). Whereas the observations show linearly rising rotation
curves out to core radii greater than $1 \kpc$ in certain dwarf galaxies
where the density is dominated by dark matter everywhere (indicating
that the dark matter has a constant density core), the simulations
predict that the collapse of collisionless particles of cold dark matter
produces cuspy halo density profiles, with a logarithmic slope
$-d\log\rho / d\log r > 1$ down to the smallest resolved radius. A
second problem is that the number of dwarf galaxies observed in the
Local Group is much smaller than the total number predicted from
numerical simulations (\cite{kkvp99}; \cite{moorea99}).

  A solution to this discrepancy has been proposed by
Spergel \& Steinhardt (2000): if the dark matter is self-interacting,
with large enough cross section to make most particles in the inner core
of a dwarf galaxy interact among themselves over a Hubble time, then an
isothermal core will be produced. A clear prediction of this hypothesis
is that when most of the particles of a halo within some radius $r_c$
have interacted, then the halo should be close to spherical inside
$r_c$, or else be supported by rotation, because the velocity dispersion
tensor should become isotropic. This paper examines the consequence of
this prediction for the inner parts of rich clusters of galaxies, where
highly magnified images of background galaxies are occasionally
observed. We will find that severe restrictions on the
collisional dark matter hypothesis are obtained.

\section{The Collisional Radius in Dwarf Galaxies and in Galaxy Clusters}

  We assume that a halo of self-interacting dark matter has an initial
density profile equal to the one for the case of collisionless dark
matter, and is thereafter modified by the effects of the collisions.
Numerical simulations of collisionless CDM models have shown that halos
have a characteristic density profile, with a logarithmic slope that
increases gradually with radius (Navarro, Frenk, \& White 1996, 1997;
Moore \etal 1999b). We define the radius $r_h$ where the logarithmic
slope is equal to 2, so that
$ | d\log\rho/d\log r | < 2$ at $r < r_h$, and 
$ | d\log\rho/d\log r | > 2$ at $r > r_h$.
The particles closest to the center will be the first ones to collide,
owing to the higher density. We define the collisional radius, $r_c$, as
the radius within which more than half the particles have interacted.
The effects of the collisions will be to change the velocity distribution
of the particles inside the collisional radius toward a Maxwellian
distribution, with constant velocity dispersion. This implies that the
density profile within the collisional radius will be altered toward
that of an isothermal sphere with finite core. The core radius produced
by the collisions can obviously not be larger than the collisional
radius, but it can be much smaller than the collisional radius if the
initial slope of the halo profile inside $r_c$ was already close to
isothermal, because the total energy needs to be conserved. Several
numerical simulations have recently been done to model this effect
(e.g., Burkert 2000, Yoshida \etal 2000, Dav\'e \etal 2001).

  In the initial density profile, the velocity dispersion should clearly
decrease toward the center at $r < r_h$: as long as the density profile
has a central power-law cusp, and the orbits are not all highly radial
near the center, then $\sigma^2(r) \propto \rho(r) r^2$. The collisions
will therefore transport heat to the colder central particles from the
hotter exterior, destroying the cusp and slowly increasing the core of
the isothermal sphere as the collisional radius increases. However, the
particles at $r> r_h$ should have a decreasing velocity dispersion with
radius in their initial configuration, so when $r_c > r_h$ heat starts
to be transported outward and the isothermal core shrinks as more
particles are slung to the outer parts of the halo (or to unbound
orbits), leading eventually to core collapse. As discussed by Spergel \&
Steinhardt (2000), the cross section should be low enough so that the core
collapse of the dark matter has not taken place in any halos up to the
present time.

  How should the collisional radius vary with the velocity dispersion
of a dark matter halo? We assume that the cross section for the elastic
collisions in the dark matter is independent of velocity, as expected
in the low energy limit when the cross section is dominated by the
s-wave contribution (e.g., Landau \& Lifshitz 1977).
Then, the rate of interaction of a particle
is proportional to the dark matter density, $\rho$, times the velocity
dispersion $\sigma$. Hence, $\rho\, \sigma\, t = {\rm constant}$, where
$t$ is the age of the halo (or the time since the last merger which
determined an initial density profile). Assuming that the core of the
halo is not larger than the collisional radius, dynamical equilibrium
implies $\rho(r_c) \propto \sigma^2/r_c^2$, and therefore,
\begin{equation}
 r_c \propto \sigma^{3/2}\, t^{1/2} ~.
\label{colprop}
\end{equation}
This implies that {\it if the core radii in dwarf galaxies are caused by
dark matter collisions within a larger collisional radius, then all the
galactic and cluster dark matter halos should have much larger
collisional radii as their velocity dispersion increases.}

  Typically, the constant density cores of dwarf galaxy halos measured
from the kinematics of the HI gas extend out to
a few kpc, and a typical velocity dispersion is $50 \kms$. As a few
examples, the rotation curves of the dwarfs DDO 154, DDO 170, and
DDO 236 yield fits for their dark matter halos with velocity
dispersion $\sigma=(28,52,45) \kms$, and core radii $(3,2.5,6) \kpc$
(\cite{cb89}; \cite{lsv90}; \cite{jc90}), with assumed distances of
$(4, 15, 1.7) \mpc$, respectively.

  If we wish to explain the sizes of these dark matter cores in dwarf
galaxies as the result of collisional dark matter, then the collisional
radii of the halos of these dwarfs must be larger than the observed
core radii, and the collisional radii in rich clusters of galaxies must
be much larger, according to (\ref{colprop}). Using the conservative
values of $r_c=2 \kpc$ and $\sigma=50 \kms$ for a typical dwarf galaxy,
and assuming that a typical rich cluster is about a third as old as a
dwarf galaxy (since massive halos have collapsed more recently than
dwarf galaxies; see Fig. 10 of Lacey \& Cole 1993), we infer that the
collisional radius of a typical rich cluster with velocity dispersion
$\sigma = 1000 \kms$ should be at least $r_c > 100 \kpc$.

  Within the collisional radius, the halo potential should be very
nearly spherical because the collisions should make the velocity
dispersion tensor of the dark matter particles isotropic (unless
the core is rapidly rotating, which is highly unlikely as will be
discussed in \S 4). This is most easily seen for a finite system,
using the tensor virial theorem: the potential energy tensor (which
reflects the shape of the mass distribution) will become diagonal
over the same timescale as the kinetic energy tensor.
The next section discusses the evidence from
gravitational lensing showing that cluster cores are elliptical in
their inner parts, focusing in particular on the example of
MS2137-23.

\section{The core of the cluster MS2137-23 is elliptical}

  Highly magnified images of background galaxies (or ``arcs'') produced
by gravitational lensing have been observed in many clusters of
galaxies. In general, models that reproduce the positions and shapes of
these images assume the presence of elliptical clumps of dark matter
centered on the most luminous galaxies in the cluster, with the
ellipticity being oriented along the same axis as the optical light.
Examples of clusters that have been modeled in this way
include A370 (\cite{kmfm93}), A2218 (\cite{kneiba95}), MS2137-23
(\cite{mfk93}), and A2390 (\cite{pierrea96}).
It should be noted that the optical isophotes of the
central cluster galaxies generally extend out to the radius where the
gravitationally lensed images are observed, where the potential is
strongly dominated by the dark matter. The regular elliptical
isophotes of the distribution of stars implies that the gravitational
potential has the same shape, and this is confirmed by the lensing
models that reproduce the positions and shapes of the multiple images
of background galaxies.
  
  We note here the intriguing fact that the isophotes of central
cluster galaxies tend to show a decrease of the ellipticity toward the
center, within radii $\lesssim 10 \kpc$ (Porter \etal 1991). This might
plausibly be an indication of the effects of self-interacting dark
matter at this small radius, making the potential more spherical;
however, other dynamical effects associated with the formation of these
galaxies from mergers
might also explain this if the dark matter is collisionless. In this
paper, we will discuss the evidence that if there is self-interacting
dark matter, the collisional radius in rich clusters of galaxies should
be smaller than $\sim 100$ kpc, leaving the question of whether there
might a smaller collisional radius for future work.

  Here, we shall focus on the cluster MS2137-23. This cluster has
several characteristics that make it particularly useful for our
purpose. First, the central region of the cluster appears to be well
relaxed as shown from both the optical image, dominated by the central
galaxy, and the X-ray emission, centered on the galaxy and with an
ellipticity and position angle similar to that of the central galaxy
(\cite{hammera97}). In clusters with substructure, the presence of
multiple mass clumps requires models of the mass distribution with many
parameters, making it difficult to constrain the ellipticity of each
mass clump. Second, a total of five gravitationally lensed images 
arising from two sources are observed in MS2137-23, providing many
constraints for the lensing model. Although redshifts for these five
images have not yet been measured, their morphologies and colors provide
strong evidence for the lensing interpretation (\cite{hammera97}).
One source produces a long, tangential arc and two other arclets,
and the second source gives rise
to a radially elongated image near the center and another arclet
(where ``arclet'' refers to images that are not magnified by very
large factors, but still show a characteristic stretching effect
due to lensing).

  The positions and relative sizes and shapes of these five
images can be reproduced in an extremely simple model: an elliptical
mass clump centered on the central galaxy, with the same ellipticity
and position angle (\cite{mfk93}; \cite{m95}).
This model needs only two free parameters for the radial density
profile (the velocity dispersion of the cluster and the core radius).
Since the positions of the five images alone already provide 6
constraints (ten coordinates of the five images minus 4 for the unknown
positions of the two sources), and in addition the relative sizes and
orientations of each image are also reproduced, this should be
considered as strong evidence that the potential of the dark matter
is elliptical, just like the stellar isophotes, and has not been
significantly circularized by dark matter collisions at the radius
where the images are observed. This radius is 15'' for the longest
tangential arc, which corresponds to 70 kpc (for
$H_0=70\kms\mpc^{-1}$). The radially elongated image is only 5'' from
the cluster center; however, if the potential became spherical only
at this small radius, this radial image would not be significantly
altered.

  Could other perturbations to the potential, arising from
substructure (which causes external shear), mimic the effect of
ellipticity if the true potential was spherical within $\sim 100$ kpc?
There are two arguments against this possibility. First, 
an external shear would be roughly constant within the region of the
multiple images, whereas an elliptical potential causes a variable
shear and convergence that depend on the density profile
(see eqs. 2 to 8 below). Second, there would be no reason why the external
shear should be aligned with the major axis of the galaxy. While
substructure is common in many clusters, the central parts of MS2137-23
appear relaxed, as discussed above.

  Although the fact that the simple elliptical potential, with constant
ellipticity as a function of radius, fits the observed positions
and shapes of the five images can already be considered as persuasive
evidence that the potential cannot be spherical within $\sim 100 \kpc$,
it will be useful to show analytically why an ellipticity is required
in a model-independent manner. We will
focus here on the radial image and its counterimage. These two images
of the same source are labeled as A1 and A5 in Mellier \etal (1993),
and in Figure 1 of Miralda-Escud\'e (1995), and as AR and A5 in
the HST image presented in Hammer \etal (1997).

\begin{figure}
\plotone{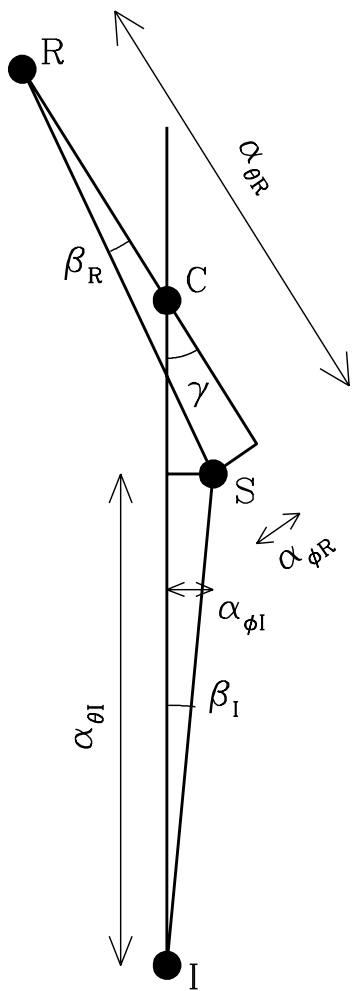}
\caption{Schematic representation of the lensing configuration
in the cluster MS2137-23 discussed in \S 3. The lens center is at $C$,
the source is located at $S$, and its two images are observed at
$R$ (which is the image on the radial critical line) and $I$. The angle
of misalignment $\gamma$ between the two images relative to the center
would be zero for a spherical potential. The radial and azimuthal
components of the deflection angles ($\alpha_{\theta}$ and $\alpha_{\phi}$)
are indicated.}
\end{figure}

  A schematic representation of the lensing of the source on the radial
caustic is shown in Figure 1, which defines the notation that will be
used here. The point labeled $C$ is the center of the cluster,
and $S$ is the position of
the source that gives rise to the radial image at $R$ and the
counterimage at $I$ (the entire lensing configuration in this system,
with the critical lines and caustics of a simple elliptical potential,
is shown in Fig. 1 of Miralda-Escud\'e 1995).
We use polar coordinates on the image
plane: $\theta$, the angular distance from the center $C$, and $\phi$,
the azimuthal angle. The light ray observed at $R$ is deflected by
an angle $\alpha_{\theta R}$ in the radial direction, and
$\alpha_{\phi R}$ in the azimuthal direction, and the same for the
light ray observed at $I$.

  The specific observed quantity that we will relate to the ellipticity
of the potential is the angle $\gamma$ of misalignment between the
images $R$ and $I$, relative to the center of the lens. In a spherical
potential, the images $R$ and $I$ should lie on a straight line passing
through $C$. The observed angle is $\gamma = 19^\circ$, indicating that
the potential is elliptical. In principle, this misalignment could
also be caused by substructure in the cluster, but this is unlikely
in view of the relaxed appearance of the cluster.

  We now relate the angle $\gamma$ to the ellipticity and the density
profile of the potential. If the ellipticity $\epsilon$ is small, 
the projected potential is adequately approximated with a quadrupole
term (e.g., \cite{m95}),
\begin{equation}
\psi(\theta,\phi) = \psi_0(\theta) - {\epsilon\over 2} \,
\psi_1(\theta) \, \cos(2\phi) ~,
\end{equation}
where
\begin{equation}
\psi_0(\theta) = \int_0^\theta d\theta ' \, \alpha_0(\theta ') ~,
\end{equation}
\begin{equation}
\alpha_0(\theta) = {2\over \theta}\, \int_0^{\theta '} d\theta ' \,
\theta ' \,  \kappa_0(\theta ') \equiv \theta \bar\kappa_0(\theta ') ~,
\end{equation}
\begin{equation}
\psi_1(\theta) = {2\over \theta^2}\,
\int_0^\theta d\theta ' \, \theta '^3 \,  \kappa_0(\theta ') ~,
\label{psi1eq}
\end{equation}
and where the surface density of the lens is
\begin{equation}
\kappa(\theta,\phi) = \kappa_0(\theta) - {\epsilon\over 2} \,
\theta\, {d\kappa_0\over d\theta} \, \cos(2\phi) ~.
\end{equation}
Here, $\kappa_0(\theta)$ is the azimuthally averaged surface density
profile, and $\bar\kappa_0(\theta)$ is the averaged surface density
within $\theta$. The deflection angle is given by the gradient of the
potential,
\begin{equation}
\alpha_{\theta} = \theta \bar\kappa_0(\theta) +
\theta \left[ \kappa_0(\theta) + {\psi_1(\theta)\over \theta^2 } \right] \,
\epsilon \cos(2\phi) ~,
\end{equation}
\begin{equation}
\alpha_{\phi} = { \psi_1(\theta)\over \theta } \, \epsilon \sin(2\phi) ~.
\end{equation}

  In the limit of a small ellipticity of the potential, the angle of
misalignment $\gamma$ is given by (using the notation in Fig. 1),
\begin{equation}
\gamma = \beta_I\, {\alpha_{\theta I}\over \theta_I - \alpha_{\theta I} }
+ \beta_R\, { \alpha_{\theta R} \over \alpha_{\theta R} - \theta_R } =
{ \alpha_{\phi I} \over  \theta_I - \alpha_{\theta I} } + 
{ \alpha_{\phi R} \over  \alpha_{\theta R} - \theta_R  } ~.
\end{equation}

  Using the condition that the rays at images $R$ and $I$ are deflected
to the same position $S$, which is simply
$\theta_i - \alpha_{\theta I} = \alpha_{\theta R} - \theta_R$
(the ellipticity introduces only second order corrections here),
we obtain
\begin{equation}
\gamma = \left[ { \psi_1(\theta_R) \over \theta_R } + 
{ \psi_1(\theta_I) \over \theta_I } \right ] \,
{\epsilon \sin (2\phi_I) \over \theta_I - \alpha_{\theta I} } ~.
\end{equation}

  We now want to find a lower limit to the ellipticity necessary to
generate the observed angle $\gamma$. For this purpose, it will be
convenient to replace the function $\psi_1(\theta)/\theta)$ by an upper
limit. Using equation (\ref{psi1eq}), we find that if the $\kappa_0$ is
constant within $\theta$, then $\psi_1(\theta)/\theta = \theta\,
\bar\kappa_0(\theta) / 2$, while in any profile where $\kappa_0$
decreases with radius, we have $\psi_1(\theta)/\theta < \theta\,
\bar\kappa_0(\theta) / 2$, because the integral of equation
(\ref{psi1eq}) weights more heavily the surface density near $\theta$
than at smaller angular radius. Therefore,
\begin{equation}
\gamma <  { \left[ \theta_R \bar\kappa_0(\theta_R) +
\theta_I \bar\kappa_0(\theta_I) \right] \, \epsilon \sin(2\phi_I) \over
 2 \, \theta_I \, \left[ 1 - \bar\kappa(\theta_I) \right] }  =
{ (1 + \theta_R / \theta_I) \, \epsilon \sin(2\phi_I) \over
 2\, \left[ 1 - \bar\kappa(\theta_I) \right] }  ~.
\end{equation}
We can now substitute the observed values $\theta_I=22\arcsec 5$
(\cite{flhc92}), and $\theta_R = 5\arcsec 2$ (\cite{hammera97}):
\begin{equation}
\gamma < 0.62\, {\epsilon \cos(2\phi_I) \over 1-\bar\kappa_0(\theta_I) } ~.
\label{glt2}
\end{equation}
To obtain a lower limit to $\epsilon$, we need to assume an upper limit
for $1-\bar\kappa(\theta_I)$. Because the two images $R$ and $I$ result
from radial (rather than tangential) magnification, there is no reason
why $\bar\kappa$ needs to be particularly close to unity at either
image. Given the relation $[\bar\kappa_0(\theta_R) - 1 ] / [1 -
\bar\kappa_0(\theta_I) ] = \theta_I /\theta_R = 4.3$, the quantity $1 -
\bar\kappa(\theta_I)$ could be very small only if the surface density
profile was very flat between the angular radii $\theta_R$ and
$\theta_I$. This is very unlikely because the velocity dispersion
implied for the cluster for an Einstein radius close to $\theta_I =
22\arcsec 5$ is already larger than $1000 \kms$ (see \cite{m95}, Figs. 8
and 9), and it would increase to a much higher value at large radius if
the slope of the density profile was much shallower than isothermal at
$\theta\sim \theta_I$.

  As a reasonable limit on how flat the $\bar\kappa$ profile could
be from $\theta_R$ to $\theta_I$, we will assume here
$\bar\kappa(\theta_R) / \bar\kappa(\theta_I) > 2$ (remember that
$\theta_I/\theta_R = 4.3$). This corresponds to
$1-\bar\kappa_0(\theta_I) > 0.16$, implying that the image $I$ is
not tangentially magnified by more than a factor 6, which is reasonable
given the length of the image $I$ (called A5 in \cite{hammera97}),
$\sim 3''$, and its axis ratio of $\sim 3$.

  With this condition, and using also $\cos(2\phi_I)\simeq 0.7$
(e.g., Mellier \etal 1993; we assume the major axis of the potential
is aligned with that of the central galaxy), and $\gamma = 0.33$, the
lower limit on the ellipticity from equation (\ref{glt2}) is
\begin{equation}
\epsilon > 0.77\, [1-\bar\kappa(\theta_I) ] \gtrsim 0.1 ~.
\end{equation}

  This is only a lower limit that we have obtained using only one
observational constraint, the misalignment of two images relative to
the center. The models that reproduce also the three images of the
other source require an ellipticity $\epsilon \simeq 0.2$. 

  There are other clusters that show little substructure in their
inner parts and are well modelled by an elliptical potential with
the major axis coinciding with that of the central galaxy: one is
A2218 (Kneib \etal 1995), which requires two clumps in the model, but
with the dominant agreeing in position and ellipticity with the
central cluster galaxy. Another is A963, which shows two
tangential arcs around the central giant elliptical (Lavery \& Henry
1988). In the case of
A963 the ellipticity is difficult to constrain because there are only
two images which could be from the same source or two different sources. 

\section{Discussion}

  The modeling of multiple images of background galaxies produced by
gravitational lensing in clusters of galaxies require elliptical models
of the mass distribution in order to reproduce their positions and
magnifications successfully (\cite{kmfm93}; \cite{mfk93}; \cite{kneiba95}).
The last section discussed the specific
example of MS2137-23, where the misalignment in the position of two
images relative to the cluster center can be used to constrain the
ellipticity in a model-independent way: the ellipticity of the dark
matter halo around the central galaxy must be greater than $0.1$ within
the image $I$, which is at $22\arcsec 5$ from the cluster center,
corresponding to a distance of $65 h^{-1} \kpc$. The fact that the
dark matter halos of galaxy clusters are elliptical within this small
radius implies that the dark matter particles have not collided over
the age of the cluster. As shown in \S 2, this also implies that the
observed cores of the dark matter halos in dwarf galaxies are too big
to have been caused by dark matter self-interaction, as proposed by
Spergel \& Steinhardt (2000).

  Further evidence supporting that cluster dark matter halos are
elliptical at radii $\sim 100$ kpc comes from the similarity with the
ellipticity of the optical isophotes of the central cluster galaxies in
both the magnitude of the ellipticity and the orientation of the major
axis (\cite{mfk93}; \cite{kmfm93}; \cite{kneiba95}). If the underlying
dark matter distribution became spherical due to the collisions, the
ellipticity of the stellar distribution would be reduced (although not
eliminated, owing to the anisotropy in the velocity dispersion tensor).
According to Hammer \etal (1997), the central galaxy in MS2137-23 has
ellipticity $\epsilon = 0.16 \pm 0.02$ beyond the radius of the radial
arc, and the best fit ellipticity for the lens model is $\epsilon =
0.18$ (see also \cite{kneiba95} for similar conclusions obtained in the
cluster A2218). We note again that the ellipticities of the optical
isophotes decline at a radius smaller than that probed by gravitational
lensing (Porter \etal 1991).

  The ellipticity of the cluster halo can be used to place an upper
limit on the interaction rate of the dark matter, in terms of the cross
section $s_x$ and mass $m_x$ of the dark matter particle. We assume here
that the collisional radius must be smaller than the distance from the
center to the long tangential arc and two other arclets
(these images are A01-A02, A2 and A4 in \cite{hammera97},
and they also require an ellipticity similar to that of the central
galaxy in the lensing models), which is about $70 \kpc$. The dark matter
density at this radius is $\rho \simeq \Sigma_{crit}/2r$, where
the critical surface density is $\Sigma_{crit}\simeq 1\gm\cm^{-2}$
for a source at $z_s=1$. Assuming also a cluster velocity dispersion
$\sigma=1000 \kms$ (roughly the minimum value required given the
Einstein radius of the cluster), and a cluster age
$t_c = 5\times 10^9$ years, we obtain the upper limit
\begin{equation}
{s_x \over m_x} < {1 \over \rho\, 2^{1/2}\sigma\, t_c } \simeq
10^{-25.5} { \cm^2 \over m_p } \simeq  0.02 {\cm^2 \over \gm} ~.
\end{equation}
For the dwarf galaxies DDO 154, DDO 170, and DDO 236 mentioned in \S 2,
with velocity dispersion $\sigma=(28,52,45) \kms$, and core radii
$(3,2.5,6) \kpc$, the time it would take for the collisional radius to
reach the value of their core radii if $s_x/m_x$ were equal to the above
upper limit is $t = (40,5,40)\times 10^{10}$ years, respectively [where
we have used the relation $t \propto \sigma^3/r_c^2$, from eq.\
(\ref{colprop}) ].

  The limit we have obtained on the self-interaction of the dark matter
also rules it out as an explanation for the low abundance of dwarf
galaxies in the Local Group, compared to the predictions of halo
satellites abundances from numerical simulations (\cite{kkvp99};
\cite{mooreb99}). In order to strike out the dark matter particles, the
satellite halos must be moving in an orbit inside the collisional
radius. For example, in the Milky Way halo (with $\sigma \simeq 150
\kms$), the collisional radius cannot be greater than about $6 \kpc$, if
$r_c < 100 \kpc$ in a cluster with $\sigma = 1000 \kms$ (where we use
the scaling $r_c \propto \sigma^{3/2}$).

  Finally, we mention three ways by which the collisional dark
matter hypothesis might still remain viable as an explanation of the
constant density cores observed in some dwarf galaxies. A first
possibility is that the presence of substructure in the mass
distribution of MS2137-23, or of other massive structures projected on
the line of sight of the cluster, introduces an external shear that
would modify the positions of the images. However, this seems unlikely
as discussed in \S 3, because elliptical models fit the observed
positions and shapes of the images remarkably well with fewer model
parameters than observational constraints, and an external shear
induces a lensing potential different than a constant ellipticity, and
would not generally be aligned with the major axis of the galaxy. 
The second possibility is that the ellipticity of the dark matter could
be supported by rotation, instead of anisotropic velocity dispersion.
However, halos formed by collisionless collapse are known to rotate very
slowly (\cite{be87}; \cite{wqsz92}), and the collisions would further
slow down the rotation of the central parts of the halo by enforcing
solid body rotation. Finally, there is the possibility that the cross
section for the dark matter interaction decreases with velocity. Here
we have assumed the cross section to be constant; if it were
proportional to $v^{-1}$ (see, e.g., Firmani \etal 2000), then the
constraints we have used here from gravitational lensing in clusters of
galaxies would allow a large enough collisional radius in dwarfs to
explain their dark matter core radii.

\acknowledgements

  I am grateful to Andy Gould, Paul Steinhardt and David Weinberg for
discussions and for their encouragement.

\end{document}